\documentclass[a4paper,11pt]{article}

\usepackage{fullpage}
\usepackage{amsmath}
\usepackage{amssymb}
\usepackage{amsthm}
\usepackage{graphicx}
\usepackage{times}
\usepackage[ruled,vlined]{algorithm2e}
\usepackage{algorithmic}
\usepackage[page]{appendix}

\newtheorem{lemma}{Lemma}
\newtheorem{theorem}{Theorem}
\newtheorem{corollary}{Corollary}
\newtheorem{observation}{Observation}

\newcommand{\rem}[1]{}

\title{Perfectly Balanced Allocation With Estimated Average Using Expected Constant Retries}

\author{Sourav Dutta, Souvik Bhattacherjee, and Ankur Narang\\ 
IBM Research, New Delhi, India \\ \{sodutta3, souvikbh, annarang\}@in.ibm.com}

\date{}

\begin{document}

\begin{titlepage}
\maketitle

\begin{abstract}

	Balanced allocation of online balls-into-bins has long been an active area of research for efficient load balancing and hashing applications. 
	There exists a large number of results in this domain for different settings, such as parallel allocations~\cite{parallel}, multi-dimensional 
	allocations~\cite{multi}, weighted balls~\cite{weight} etc. For sequential multi-choice allocation, where $m$ balls are thrown into $n$ bins with each ball 
	choosing $d$ (constant) bins independently uniformly at random, the maximum load of a bin is $O(\log \log n) + m/n$ with high probability~\cite{heavily_load}. 
	This offers the current best known allocation scheme. However, for $d = \Theta(\log n)$, the gap reduces to $O(1)$~\cite{soda08}. 
	A similar constant gap bound has been established for parallel allocations with $O(\log ^*n)$ communication rounds~\cite{lenzen}.

	In this paper we propose a novel multi-choice allocation algorithm, \emph{Improved D-choice with Estimated Average} ($IDEA$) achieving a constant 
	gap with a high probability for the sequential single-dimensional online allocation problem with constant $d$. We achieve a maximum load of 
	$\lceil m/n \rceil$ with high probability for constant $d$ choice scheme with \emph{expected} constant number of retries or rounds per ball. We also 
	show that the bound holds even for an arbitrary large number of balls, $m>>n$. Further, we generalize this result to (i)~the weighted case, where  
	balls have weights drawn from an arbitrary weight distribution with finite variance, (ii)~multi-dimensional setting, where balls have $D$ dimensions 
	with $f$ randomly and uniformly chosen filled dimension for $m=n$, and (iii)~the parallel case, where $n$ balls arrive and are placed parallely in the 
	bins. We show that the gap in these case is also a constant w.h.p. (independent of $m$) for constant value of $d$ with expected constant number of retries per ball.
  
\end{abstract}
\end{titlepage}

\section{Introduction}
\label{sec:intro}

A central research area in the domain of randomized algorithms is the \emph{occupancy problem} for \emph{balls-into-bins} processes~\cite{azar, perfect, heavily_load, lenzen, mitzen}. The framework of the problem involves the analysis of the online 
allocation, wherein a set of independent balls is to be assigned to a set of bins. The occupancy problem helps to model several realistic problems into a formal mathematical structure, and hence opens an active area of work in probability theory 
as well as in computer science.

In the classical ``balls-into-bins'' problem, $m$ balls are sequentially thrown into $n$ bins, where each ball is placed into one of the bins independently and uniformly at random (i.u.r.). The natural question then is to analyze the maximum load in any of the bins. Mapping the problem to the application domain, we may consider the balls to be jobs or tasks and the bins to be servers. The problem then reduces to scheduling the jobs with balanced load allocations among the servers.

Probably one of the earliest applications of randomized load balancing is in the context of $hashing$. For the chaining method during hash clash, the length of the lists in the hash buckets are a measure of the retrieval complexity. For a uniform 
hash function, the length of the lists follow the same distribution as the number of balls in a bin in this case.

The advent of parallel and distributed systems required efficient online load balancing among the servers to improve the 
throughput of the system. Dependence on a centralized environment for uniform load balancing is highly undesirable for 
such systems due to high communication complexity. With the introduction of the \textit{Cloud Computing} paradigm, the placement of virtual machines (VMs)
on servers provided a new dimension to the applicability of the randomized balanced allocation study.

Other applications such as the design of Multimedia or Data Servers use disk arrays where a data unit is partitioned and stored in a distributed fashion. These applications demand even (balanced) access of the disks on retrieval~\cite{sanders} and Karp in~\cite{karp} discusses applications in video-on-demand (termed \emph{k-orientability}~\cite{perfect}). The balls into bins problem accurately describes these applications only when the balls have uniform weights. 
Other applications assume the loads to be of different weights to model its various dimensions.

This paper tackles the problem of sequential online allocation of balls into bins. Assuming we have $n$ bins and 
$m$ balls arriving one at a time are to be thrown into these bins, the problem is to devise an efficient algorithm such 
that the allocation of the balls is nearly balanced among all the bins. In formal terms, the 
load in each of the bins should be as close to the average, ($m/n$) as possible. We initially study the case of single-dimensional 
sequential placement of uniform weighted balls into bins problem and then extend it for the general weighted case. Finally we also observe 
that $IDEA$ provides the same result w.h.p. for multi-dimensional balls-into-bins problem for $m=n$.

In this context, we define \emph{Gap} to be the difference between the heaviest loaded bin and the average load. The currently best known algorithm bounds $Gap$ to $O(\log \log n)$ with high probability using the symmetric \emph{d-choice} placement strategy~\cite{azar, mitzen}. In the d-choice method, each ball selects $d$ bins i.u.r. among the $n$ bins and is allocated to the least loaded bin among them. It is well-known that if $d = \Theta(\log n)$ choice, the gap is $O(1)$~\cite{soda08}.

In this paper we propose a novel algorithm, \emph{Improved D-choice with Estimated Average}, ($IDEA$) for efficient placement of the balls in the bins. We prove that this technique provides a \emph{\textbf{constant}} $Gap$ with high probability (w.h.p.) even when $d$ is kept constant, albeit with an expected constant number of retries or rounds per ball. We further extend the result to show that the guarantee also holds true for the heavily loaded case, i.e. $m>>n$ w.h.p. Our technique is different from the typical greedy $d$-choice process in that it places the ball in the bin that has load equal-to or lower than the \textit{estimated average} of that bin. Using \textit{expected} constant number of retries such a bin can be found for each ball and hence the load in each bin tends towards the estimated average which also tends towards the actual average, resulting in constant upper bound on the gap. Our strategy is also different from the typical asymmetric strategy~\cite{vocking-tree} where in case of tie over the load, the leftmost bin gets the ball. Our result can have profound implication both theoretically and practically on the online load balancing algorithms. 

The outline of the paper is as follows: Section~\ref{sec:rel} presents an introduction to the known works and results in this domain. In Section~\ref{sec:algo} we propose the detailed outline of the $IDEA$ algorithm for allocating the balls into the bins. Section~\ref{sec:proof} provides the theoretical proof for bounding the $Gap$ to a constant quantity with high probability. Section~\ref{sec:disc} provides insights into the execution of the $IDEA$ algorithm. 
Section~\ref{subsec:ext_weight} depicts its extension for the general weighted balls case, Section~\ref{subsec:ext_multi} exhibits similar results for the multi-dimensional scenario, and Section~\ref{subsec:parallel} proposes the protocol for achieving the same results for the parallel scenario. Finally, Section~\ref{sec:conc} concludes the paper.

\section{Related Work}
\label{sec:rel}

The study of ``balls-into-bins'' problem dates back to the study of hashing by Gonnet. He showed that when $n$ 
balls are thrown into $n$ bins i.u.r., the fullest bin has an expected load of $(1+o(1))\log n / \log \log n$~\cite{gonnet}.
The maximum loaded bin in this approach was shown to be $O(\log n/ \log\log n)$ w.h.p.~\cite{ranjan}. It was also shown 
that for $m \geq n\log n$ balls, a bin can have a maximum load of $m/n + \Theta(\sqrt{m\log n/ n})$.

Azar et al.~\cite{azar} showed that if the balls chose sequentially from $d \geq 2$ bins i.u.r. (called \textit{Greedy[d]} algorithm) and greedily selected the bin currently with the lowest load, the $Gap$ could be bounded by $O(\log \log n/\log d)$ w.h.p. However, the solution worked only for the case when $m=n$. They also showed that the bound is stochastically optimal, i.e. any other greedy approach using the placement information of the previous balls to place the current ball majorizes to their approach. However, if the alternatives are drawn from separate groups with different rules for tie breaking, it results in different allocations.~\cite{vocking-tree} presents such an \textit{asymmetric} strategy and using witness tree based analysis proves that this leads to an improvement in the load balance to $O(\frac{\log \log(n)}{d\log(\phi_d)})$ w.h.p. where, $\phi_2$ is the golden ratio and $\phi_d$ is a simple generalization. Our algorithm is different from both these techniques in that it uses the \textit{estimated gap} as the criterion for choosing the bin and makes potentially multiple retries, where in each retry $d$ bins are chosen i.u.r.

For the heavily loaded case, $m>>n$, the bound of $O(\log \log n/\log d)$ w.h.p. was later proven in~\cite{heavily_load} using sophisticated techniques in two main high level steps. In the first step, they show that when the number of balls is polynomially bounded by the number of bins the gap can be bounded by $O(\ln\ln(n))$, using the concept of layered induction and some additional tricks. In particular, they consider the entire distribution of the bins in the analysis (while in typical $m = O(n)$ case the bins with load smaller than the average could be ignored). In the second step, they extend this result to general $m >> n$ case, by showing that the multiple-choice processes are fundamentally different from the classical single-choice process in that they have \textit{short memory}. This property states that given some initial configuration with gap $\Delta$, after adding $poly(n)$ more balls the initial configuration is \textit{forgotten}. The proof of the short memory property is done by analyzing the mixing time of the underlying Markov chain describing the load distribution of the bins. The study of the mixing time is via a new variant of the coupling method (called \textit{neighboring coupling}).
It was also shown that when $d = \Theta(\log n)$ the gap becomes $O(1)$~\cite{soda08}.

Cole et al.~\cite{routing-cole} showed that the two-choice paradigm can be applied effectively in a different context, namely, that of routing virtual circuits in interconnection networks with low congestion. They showed how to incorporate the two-choice approach to a well-studied paradigm due to Valiant for routing virtual circuits to achieve significantly lower congestion.

Kunal et.al.~\cite{kunal-weighted} prove that for weighted balls (weight distribution with finite fourth moment) and $m >> n$, the expected gap is independent of the number of balls and is less than $n^c$, where $c$ depends on the weight distribution. They first prove the weak gap theorem which says that w.h.p $Gap(t) < t^{2/3}$. Since in the weighted case the $d$ choice process is not dominated by the one choice process, they prove the weak gap theorem via a potential function argument. Then, the \textit{short memory theorem} is proved.  While in~\cite{heavily_load} the short memory theorem is proven via coupling,~\cite{kunal-weighted} uses similar coupling arguments but defines a different distance function and use a sophisticated argument to show that the coupling converges.

The $(1+\beta)$-choice scheme~\cite{beta} proved that if a ball chooses with $\beta \in (0,1)$ probability the least loaded bin of $d=2$ randomly chosen bin, and otherwise i.u.r. a single bin, the $Gap$ becomes independent of $m$ and is given by $O(\log n / \beta)$. 

In the parallel setting,~\cite{lenzen} showed that a constant bound on the gap is possible with $O(\log^* n)$ communication rounds. Adler et.al.~\cite{parallel} consider parallel balls and bins with multiple rounds. They present analysis for $O(\frac{\log\log(n)}{\log(d)})$ bound on the gap (for $m = O(n)$) using $O(\frac{\log\log(n)}{\log(d)} + O(d))$ rounds of communication.

For offline balls-into-bins problem, using maximum flow computations it was shown that the maximum load of a bin w.h.p. is $\lceil m/n \rceil +1$.~\cite{perfect} showed that for $m>cn\log n$ balls, where $c$ is a sufficiently large constant, a perfect distribution of the balls was possible w.h.p.  However, no such similar result is found in the literature for the online sequential case for constant $d$ choice.

Mitzenmacher et. al. in~\cite{multi} addresses both the single choice and d-choice paradigm for multidimensional balls and
bins under the assumption that the balls are uniform D-dimensional (0, 1) vectors, where each ball has exactly $f$ populated
dimensions. They show that the gap for multidimensional balls and bins, using the two-choice process, is bounded by
O(log log(nD)). We provide a better bound of $O(1)$ w.h.p. for $m=n$ case.

In this paper, we study a novel online sequential allocation algorithm for balls-into-bins based on a constant \emph{d-choice} strategy and prove a 
constant gap bound both for $m=n$ and the heavily loaded case $m >> n$ along with for the general weighted balls and multi-dimensional scenario.

\section{The $IDEA$ Algorithm}
\label{sec:algo}

In this section we discuss the execution of the \emph{Improved D-choice with Estimated Average} ($IDEA$) algorithm. We consider 
there are $n$ bins and $m$ balls which arrive in an online fashion. We initially assume that the balls are of uniform weights and are numbered according to 
the order of their arrival. In hashing applications, the number of the balls based on their arrival order plays no role in assisting 
better or faster retrieval. Hence, this assumption does not decrease the complexity of the problem at hand. Later we also provide a blueprint 
of the case when such a numbering of the balls in not allowed and the weighted balls case with the weights of the balls drawn from an arbitrary 
distribution with finite variance. 

\begin{algorithm}[ht]
\begin{center}
	\caption{IDEA Algorithm}
	\label{algo:idea}
	\begin{algorithmic}
		\REQUIRE Number of bins ($n$), Number of balls ($m$) and Maximum iteration ($\gamma$)
		\ENSURE Balanced Allocation of Balls-into-Bins

		\medskip

		\FORALL{bin $B_i$, $i \in [1,n]$}
			\STATE Initialize the load, $L_{B_i}$ and estimated average, $\hat{A_{B_i}}$ to $0$
		\ENDFOR
		
		\FORALL{ball $b_j$, $j \in [1,m]$}
			\STATE $loop \gets 0$
			\WHILE{$loop \leq \gamma$}
				\STATE Choose $d$ bins, $C = \{Bin_1, Bin_2, \cdots Bin_d\}$  i.u.r. from the $n$ bins
				\IF{set $C$ contains at least one bin with negative or zero estimated gap, $\hat{Gap_{Bin_i}} = L_{Bin_i} - \hat{A_{Bin_i}}$}
					\STATE Break \textbf{while}
				\ENDIF
				\STATE $loop \gets loop + 1$
			\ENDWHILE

			\STATE Place ball $b_j$ in the bin, $B \in C$ having the lowest estimated gap, $\hat{Gap_B}$
			\STATE $L_B \gets L_B + 1$

			\FORALL{bins, $Bin_i \in C$}

				\IF{$\hat{A_{Bin_i}} > \lceil j/n \rceil$}
					\STATE $flag \gets 1$
				\ELSE
					\STATE $flag \gets 0$
				\ENDIF

				\IF{$flag = 0$}
					\STATE $\hat{A_{Bin_i}} \gets \hat{A_{Bin_i}} + 1/d$
				\ENDIF
			\ENDFOR
		\ENDFOR
	\end{algorithmic}
\end{center}
\end{algorithm}

Given each bin has an accurate knowledge of the average number of balls in the system, $m/n$ it is easy to distribute the balls so as 
to obtain a perfectly balanced allocation. $IDEA$ operates on the above principle, where each bin independently calculates a fairly good 
estimate of the current average number of balls in the system. Each bin is then loaded nearly equal to its estimated average 
value. In the remainder of this section we show how each bin independently estimates its average which we later prove, 
with a high probability, to be very close to the actual average, $m/n$. We also show that each bin is then loaded close to its 
estimated average value, giving a maximum load of $\lceil m/n \rceil$ with a constant gap allocation w.h.p.

The $IDEA$ algorithm initially works as in the d-choice algorithm. On arrival of a ball $b_j$, it i.u.r. chooses $d$ bins ($d$ is constant) as 
its possible candidates for placement. Each bin, $B_i, i \in [1,n]$ is characterised by two parameters: (i)~\emph{Current Load}, $L_i^j$, 
and (ii)~\emph{Current Estimated Average}, $\hat{A_i^j}$. For each bin we define its \emph{estimated gap}, $\hat{Gap_i^j}$ as the difference 
between its current load and its current estimated average. Formally, $\hat{Gap_i^j} = L_i^j - \hat{A_i^j}$.

The ball $b_j$ is then allocated to the bin having the lowest value of $\hat{Gap_i^j}$ among the $d$ chosen bins. 
Given the definition of $Gap$ (in Section~\ref{sec:intro}) we would like to place the ball in a bin with \emph{negative} 
or \emph{zero} $\hat{Gap}$. This would ensure that the loads in the bins be close to their estimated average values and thus lead 
to a lower $Gap$. Hence, if in the $d$ choice a ball selects no bin with negative or zero $\hat{Gap_i}$, it re-chooses 
its candidate $d$ bins. To boost the probability of a ball choosing a bin having such $\hat{Gap_i}$, this 
re-choosing will be carried out $\gamma$ times, where $\gamma$ will later be shown to be approximately a constant. 

The current estimated average for each of the $d$ bins finally selected by the ball is then incremented by 
$1/d$. In the next paragraph we discuss the selection of such an increment value. We intuitively argue that for 
each bin if $\hat{A_i}$ is finally close to the actual average ($m/n$) w.h.p., and its load $L_i$ is nearly equal to its estimated average, 
the overall $Gap$ in the system will be minimized and the maximum load of a bin will be $\lceil m/n \rceil$.  The pseudo-code of $IDEA$ algorithm 
is shown in Algorithm~\ref{algo:idea}.

The probability that a bin is chosen by a ball in its $d$ choice is given by $d/n$. So when $n$ balls arrive a bin will be chosen $d$ times 
on expectation. For each such choice the estimated average of the bin is incremented by $1/d$ (Algorithm~\ref{algo:idea}). Hence, its 
final estimated average will be $1$, which is indeed the actual average of the system. However, from Lemma\ref{lem:ping} we observe that a bin might 
be chosen $d \log n$ times or lesser w.h.p. Since we increase the estimated average by $1/d$, 
the estimated average may increase beyond $1$ in such cases. Hence the estimated average of a bin may be greater that $1$ in two situations: \\
(i)~Not more than $n$ balls have arrived, but the bin has been chosen close to $d \log n$ times, or \\
(ii)~More than $n$ balls have arrived. \\
For case (i), the estimated average of the bin should still remain $1$, while in the other case, the estimated average should be 
increased as usual. It is here that the numbering of the balls come into effect. If the estimated average of a bin goes beyond $1$ 
and the next ball which selects this bin has a number less than $n$, the bin knows that it may be chosen $d \log n$ times and hence 
refrains from increasing its estimated average until a ball with number more than $n$ selects it. Similarly when the estimated average 
of a bin increases beyond $\alpha, \alpha \in \mathbb{N}$, it checks if the next ball selecting it has a number greater than $\alpha n$. Thus the balls 
communicate their numbers as well while choosing the $d$ candidate bins.

However in the scenario where numbering of the balls is forbidden, to differentiate between the two cases, we use the sampling technique among the bins. 
A bin with estimated average just above $\alpha$, in this case chooses $\log n$ bins i.u.r. and communicates with them for their estimated average. 
If the average of the estimated averages of the sampled bins is less than $1$, the bin comprehends that case (i) 
has happened, i.e., it is receiving more than $d$ balls out of $n$ balls and thus refrains from increasing its estimated average. However, if the average of the estimated 
averages are $1$, the bin decides that more than $\alpha n$ balls are arriving and increases its estimated 
value as usual. The probability that the error in the sampled average is greater than $\epsilon$, a small constant, is given by $\frac{1}{n}$ for \emph{constant} 
number of samples when $m>n\log n$ and by $\log n$ sampled choice for $m<n\log n$ scenario (\emph{sampling theorem}). 
Hence w.h.p. of $1-\frac{2}{n}$ we obtain the right decision for each bin. In Appendix~\ref{sec:sampling} we discuss in detail 
the proof for this claim, and also show that the total number of such sampling done is less than communication done if $d=\log n$. More intelligent sampling 
methods as that of \emph{Reservoir Sampling}~\cite{reservoir}, \emph{Subset-Sum Sampling}~\cite{subset1,subset} or a combination of \emph{Sampling} and \emph{Sketching}~\cite{sketch1,sketch} 
may be used to obtain a better estimates. The study and effects of such methods are not discussed as a part of this paper.

Hence, we find that $IDEA$ dynamically adapts its estimated average to be closer to the actual average of the system. In either case, the estimated average of a bin is increased by at most $1$ for every 
$n$ balls.

\section{Theoretical Framework}
\label{sec:proof}

In this section, we provide a theoretical proof of the constant gap performance of the $IDEA$ algorithm. First, we 
bound the number of balls that may select each bin. We then establish that each 
ball in the $IDEA$ algorithm chooses at least one bin having negative $\hat{Gap}$ with a high probability, which makes the 
load of each bin converge to its estimated average value. Finally, we bound the $Gap$ of the system to a 
constant value w.h.p. We assume $m$ balls to arrive in an online fashion and there are $n$ bins.

\begin{lemma}
\label{lem:ping}
	If each ball chooses $d$ bins i.u.r. out of $n$ bins, each bin is chosen by $\frac{md}{n}$ balls on expectation, 
	and by at most $\frac{md}{n} \log n$ balls with high probability.
\end{lemma}
\begin{proof}
	Define $Y_1, Y_2, \cdots Y_m$ to be indicator random variables corresponding to balls $b_1, b_2, \cdots, b_m$ respectively. 
	Let $Y_i = 1$ represent the event that the ball $b_i$ chose bin $B$ as one of its $d$ candidate bins, otherwise $Y_i = 0$, $\forall 
	i \in [1,m]$. Since the balls choose $d$ bins i.u.r., the probability that bin $B$ is chosen among the $d$ bins, or $\Pr (Y_i) = 1$, 
	is given by $d/n$. Let $X$ be a random variable depicting the number of balls that chose $B$ among its $d$ candidate bins. 
	Hence, $X = \sum_{i=1}^m Y_i$. The expected value of $X$ is,
	{\small{
	\begin{align}
	\label{eq:exp_X}
		E[X] = E[\sum_{i=1}^m Y_i] = \sum_{i=1}^m E[Y_i] = \sum_{i=1}^m \frac{d}{n} = \frac{md}{n} \qquad \qquad \text{[By Linearity of Expectation]}
	\end{align}
	}}
	Applying Chernoff's bound on $X$ we obtain, 
	{\small{
	\begin{align}
	\label{eq:X_bound}
		&P(X > (1+\delta)E[X] ) < \frac{e^\delta}{\left(1+\delta \right)^{\left(1+\delta \right)}} \nonumber \\
		&\therefore P(X > (1+\delta)\frac{md}{n} ) < \frac{e^\delta}{\left(1+\delta \right)^{\left(1+\delta \right)}} \nonumber \\
		&\text{Substituting $\delta = \log n - 1$ we have,} \nonumber \\
		&P(X > \frac{md}{n}\log n ) < \frac{e^{\log n -1}}{(\log n) ^{\log n}} = \frac{n}{e (\log n)^{\log n}} \\		
		&\text{Let $y = (\log n) ^{\log n}$. Hence, $\log y = \log n \log \log n$. We have,} \nonumber \\
                &\Rightarrow \log (y/n) = \log n \left(\log \log n - 1\right) = \log n \left(\log \log n - \log \log e^e\right) \nonumber \\
		&\text{For large values of $n$, $\log \log (n/e^e) \geq 1$, giving $\log (y/n) \geq \log n$. Therefore, we have $y > n^2$.} \nonumber \\
		&\text{Substituting in Eq.~\eqref{eq:X_bound},} \nonumber \\
	\label{eq:X_prob}
		&P(X > \frac{md}{n}\log n ) < \frac{1}{en}
	\end{align}
	}}
	Hence, bin $B$ is chosen by at most $\frac{md}{n} \log n$ balls with a high probability of $1 - \frac{1}{en}.$
\end{proof}

\begin{lemma}
\label{lem:av}
	At any iteration, the estimated average of each bin is approximately equal to the current average with high probability.
\end{lemma}
\begin{proof}
	We assume here that $Z$ balls have already arrived and have been placed among the $n$ bins. The number of balls that chose bin $B$ among its $d$ candidates is 
	$\frac{Zd}{n}$ on expectation, since each bin can be chosen by a ball with a probability of $\frac{d}{n}$. The number of such balls is also bounded 
	by $\frac{Zd}{n}\log n$ with high probability (by Lemma~\ref{lem:ping}). However, a bin does not increment its estimated average by more than $d$ 
	times for every $n$ balls. For each choice the bin $B$ increases its estimated average by $\frac{1}{d}$. Hence the current value of $\hat{A_B}$ is given by,
	{\small{
	\begin{align}
		\hat{A_B} = \frac{Zd}{n} \cdot \frac{1}{d} = \frac{Z}{n} \text{  , which is the current average.} \nonumber
	\end{align}
	}}
	Hence, the estimated average $\hat{A}$ of any bin is nearly equal to the actual average w.h.p.
\end{proof}

\begin{observation}
\label{obs:var}
	The variance of the estimated average of a bin $B$ for $n$ balls is,
	{\small{
	\begin{align}
		Var[\hat{A_B}] &= Var[\frac{X}{d}] = Var[\frac{1}{d}.\sum_{i=1}^n Y_i] = \frac{1}{d^2} \sum_{i=1}^n Var[Y_i] \nonumber \\
		&= \frac{1}{d^2}.n\frac{d}{n}(1-\frac{d}{n}) = \frac{1}{d} - \frac{1}{n} \qquad \qquad \qquad \text{[From Lemma~\ref{lem:ping}]} \nonumber
	\end{align}
	}}
\end{observation}

\begin{lemma}
\label{lem:gap}
	The amortized sum of the estimated gap, $\hat{Gap}$ over all the bins is zero after every $n$ balls.
\end{lemma}
\begin{proof}
	Each ball chooses $d$ candidate bins i.u.r. and is finally allocated to the bin having the least estimated gap. Hence for all 
	the $d$ chosen bins, their estimated average is increased by $1/d$.  The bin which receives the ball witness an increase in its actual 
	load by $1$. Hence, overall its estimated gap increases by $1-1/d$. However, for the remaining $d-1$ bins their loads remain the same, 
	and thus their estimated gap decreases by $1/d$. Hence the overall change in estimated gap over the $d$ chosen bins is $1-1/d + (d-1)(-1/d) = 0$.
	Initially, since the sum of the estimated gaps of the bins was $0$, the lemma holds.

	Considering a batch of $n$ balls arriving in the system, a bin may be selected more than $d$ times (Lemma~\ref{lem:ping}). In such case, 
	the bin samples other bins for their current estimated average value, and depending on it may or may not increase its estimated average 
	as discussed in Section~\ref{sec:algo}. 
	As such the change in the overall estimated gaps in this round will not add up to $0$. Such a scenario occurs when a bin is selected more 
	than $d$ times in the batch of $n$ balls. Such a bin may not increase its estimated average, and $IDEA$ experiences a positive change in 
	the overall estimated gap of the system for such a round.

	However, it can be observed that for a batch of $n$ balls, the total number of bins that are selected by the balls is exactly $nd$.
	Since we consider a bin to have been selected more than $d$ times, there exists at least one bin which was selected less than $d$ times. 
	Assume a \emph{bank} to exist, which loans a unit credit to the bin, selected more than $d$ times for $n$ balls, per extra selection. 
	If such a bin is selected $d+c$ times over a period of $n$ incoming balls, the total credit units in the bank is exactly $c$. However, 
	since the number of selections are fixed, the total \emph{holes} in the system will also be exactly be equal to $c$. \emph{Hole} in a bin 
	refers to the difference of $d$ and the number of times the bin has been selected by $n$ balls, for bins selected less than $d$ times. 
	Each such bin can be considered to have extra unit credit points per hole, which it returns to the bank after $n$ balls have been 
	allocated to the system. Since the number of credits in the bank is exactly equal to the number of extra credits held by the bins in the 
	system, after $n$ balls the total credit points of the bank will be $0$.

	It can easily be observed that the total credits in the system is always a non-negative quantity. Since the bins are chosen by the balls 
	i.u.r., all the bins are selected nearly the same number of times over a period of $n$ balls, no bins tends to accumulate a large quantity of 
	extra credits that it always keeps returning to the bank. This factor helps to maintain the estimated average of each bin close to the actual 
	average of the system. Hence, combining both the settings, we prove that on an \emph{amortized} notion, the sum of the estimated gap in all 
	the bins is $0$ after every $n$ balls.
\end{proof}

\begin{corollary}
\label{cor:gap_zero}
	The sum of the \emph{estimated gap} over all bins is zero for arbitrary small number of balls allocated in the system.
\end{corollary}
\begin{proof}
	Let the number of balls being allocated in the system be a function of $n$, $f(n)$. Given the constraint that the value of $f(n)$ is not a constant, 
	the arguments of Lemma~\ref{lem:gap} still holds true. Consider, $f(n) = n^{\epsilon}$, where $\epsilon$ is arbitrary small respecting the 
	constraint that $f(n)$ is not a constant. Thus, the sum of the estimated gap in the system is $0$ after $f(n)$ balls have been allocated to the 
	bins.
\end{proof}

\begin{lemma}
\label{lem:negbins}
	The number of bins having a \emph{zero} or \emph{negative} estimated gap, $\hat{Gap}$ is $\Theta(n)$.
\end{lemma}
\begin{proof}
	In Lemma~\ref{lem:gap} and Cor.~\ref{cor:gap_zero}, we show that the sum of the estimated gap of the bins is $0$ even when 
	arbitrarily small number of balls are allocated to the bins. As such the number of bins with positive estimated gap cannot 
	increase by more than $n^{\epsilon}$.
	
	Let there be $\alpha$ bins with positive $\hat{Gap}$, $\beta$ bins with negative estimated gap, and $\theta$ bins having $0$ 
	estimated gap. Hence, $\alpha + \beta + \theta = n$. We would like to establish a lower bound on $\beta +\theta$. 
	In order to have minimum number of bins with negative or zero $\hat{Gap}$, the value of the gap should be minimum for bins with 
        a positive gap and maximum for bins with a negative gap. The minimum positive estimated gap for a bin is
        $Z(1-\frac{d-1}{d})$ when $Z(d-1)$ balls have arrived in the system, of which only $Z$ balls have been committed into the bin.
        The maximum negative estimated average that a bin may have in this case is $-\frac{Z(d-1)}{d}$. Hence,
	{\small{
	\begin{align}
		&\alpha.(Z(1-\frac{d-1}{d})) + \beta.(-\frac{Z(d-1)}{d}) + \theta .0 = 0 \qquad \qquad \qquad \text{[From Lemma~\ref{lem:gap}]} \nonumber \\
		&\therefore \alpha = \beta(d-1) \nonumber
	\end{align}
	}}
	As $\alpha + \beta + \theta = n$, we have $d\beta + \theta = n$. Hence, the number of bins with zero or negative $\hat{Gap}$ is $\Theta(n)$.

	For each round of $f(n)$ balls, the number of bins with zero or negative estimated gap may decrease by $f(n)$. Consider that in round $k$, the number 
	of bins with zero or negative gap is $N(c_k)$. In the $(k+1)^{th}$ round, the number of such bins may become $N(c_k) - f(n)$.  However, as $f(n)$ is 
	considered to be very small, in the order notation the number of such bins still remains $\Theta(n)$. We contradict the existence of any additive 
	influence of $f(n)$ per round by the argument of amortized analysis in the above lemma and its corresponding corollary.
\end{proof}

\begin{lemma}
\label{lem:negchoice}
	Each ball chooses at least one bin having negative estimated gap among its $d$ choices w.h.p. in $\gamma$ rounds.
\end{lemma}
\begin{proof}
	Each ball selects independently and uniformly at random $d$ candidate bins for its placement among 
	the $n$ bins. Hence the probability that bin $B_i$ is chosen as a candidate for ball $b_j$ is, 
	$P_i^j = \binom{n-1}{d-1} / \binom{n}{d} = \frac{d}{n}$. Let there be $c$ bins with zero or negative $\hat{Gap}$. The 
	probability that neither of these bins are selected as candidate by a ball $ = \binom{n-c}{d} / 
	\binom{n}{d}$. The ball may re-select its candidates at most $\gamma$ times. Therefore, the probability 
	that neither of the $c$ bins are selected in any of the $\gamma$ tries $=\left(\binom{n-c}{d} / 
	\binom{n}{d}\right)^\gamma$. Hence the probability that at least one bin with negative $\hat{Gap}$ 
	is selected in the $\gamma$ iteration is given by,
	{\small{
	\begin{align}
	\label{eq:one_sel}
		P(\text{at least one selected}) = 1 - \left(\frac{\binom{n-c}{d}}{\binom{n}{d}}\right)^\gamma \approx 1 - \frac{1}{2^{d\gamma}} \qquad 
		\text{[Assuming $c=n/2$ from Lemma~\ref{lem:negbins}]}
	\end{align}
	}}
	For $d=2$ and $\gamma=2$, we obtain a probability of around $0.94$. However, with $\gamma = \log n$, the probability becomes nearly $1-\frac{1}{n}$. Further, we can show that approximately constant number of retries suffice.
	
	Let the number of bins with positive gap at any point of time be $n^{1-\epsilon}$, where $0 \leq \epsilon \leq 1$. The probability $P_{bneg}$ with which a bin 
	with a zero or negative gap is chosen in $\gamma$ iterations is given by,
	{\small{
	\begin{align}
	 	&P_{bneg} = 1 - \left(\frac{\binom{n^{1-\epsilon}}{d}}{\binom{n}{d}}\right)^{\gamma} \nonumber
	\end{align}
	}}
	For a zero or a negative bin to be chosen with a high probability, we need $P_{bneg} \ge 1 - \frac{1}{n^{\phi}}$, where $\phi > 0$. Hence for $\left(1 - (\frac{n^{1-\epsilon}}{n})^{d\gamma}\right) > 1 - \frac{1}{n^{\phi}}$. Thus, $\gamma > \frac{\phi}{d\epsilon}$.
	Hence, at least one such bin is chosen by each ball in approximately 
	constant $\gamma$ re-polls or rounds per ball w.h.p.
\end{proof}

In the next lemma, we show that in practice only a couple of retries are needed to get a bin with zero or negative estimated gap.

\begin{lemma}
\label{lem:constgamma}
	The expected number of rounds, $\gamma$ per ball to find a bin with zero or negative estimated gap is constant.
\end{lemma}
\begin{proof}
	Let $p_i$ denote the probability that we find a zero or a negative bin at iteration $i$. Therefore, we have 
	{\small{
	\begin{align}
		p_i &= \left(\prod_{1}^{i-1} P_{pos}\right)\cdot P_{neg} = \prod_{1}^{i-1} \frac{1}{2^d} \cdot \left(1-\frac{1}{2^d}\right) = \frac{2^d - 1}{2^{id}} \nonumber
	\end{align}
	}}
	where $P_{pos}$ is the probability of selecting a bin with a positive estimated gap and $P_{neg}$ is the probability of selecting a bin with a zero or negative gap.
	The expected number of rounds per ball, $\gamma$ to find a zero or a negative gap is given by,
	{\small{
	\begin{align}
	\label{eq:exp}
		&E[\gamma] = \sum ip_{i} = \left(2^d-1\right)\sum_{i=1}^{\frac{\log{n}}{d}}\frac{i}{2^{id}}
	\end{align}
	Let,
	\begin{align}
	\label{eq:si}
		&S(i) = \sum_{i=1}^{\frac{\log{n}}{d}}\frac{i}{2^{id}} \\
	\label{eq:si+1}
		&\therefore \frac{S(i)}{2^d} = \sum_{i=1}^{\frac{\log{n}}{d}}\frac{i}{2^{(i+1)d}}
	\end{align}
	Subtracting Eq.~\eqref{eq:si+1} from Eq.~\eqref{eq:si}, we have
	\begin{align}
	\label{eq:sum}
		&\left(1-\frac{1}{2^d}\right)S(i) = \frac{1}{2^d} - \frac{\log{n}}{d2^{1+\log{n}}} + \frac{1}{2^d\left(2^d-1\right)} \nonumber \\
		&\therefore S(i) \approx \frac{2^d}{\left(2^d-1\right)^2}
	\end{align}
	Substituting Eq.~\eqref{eq:sum} in Eq.~\eqref{eq:exp}, we have
	\begin{align}
	 	E[\gamma] &\approx 1 + \frac{1}{2^d-1}\\ \nonumber
	 	\Rightarrow E[\gamma] &< 2
	\end{align}
	}} 
	Given the number of bins having negative of zero estimated gap to always remain $\Theta(n)$, the number of retries per balls remains constant 
	throughout the execution of the $IDEA$ algorithm.
\end{proof}

\begin{lemma}
\label{lem:load}
	The load of each bin tends to its estimated average.
\end{lemma}
\begin{proof}
	$IDEA$ places each ball into a bin with zero or negative $\hat{Gap}$, with high probability $1 - \frac{1}{n^\phi}$ (Lemma~\ref{lem:negchoice}) using $\gamma$ retries. When a ball is placed in a bin, its $\hat{Gap}$ increases. Thus, the probability that this bin will again get a ball lowers. On the other hand, the bins that had been chosen but the ball was not placed in them have a decrease in their estimated gap. Hence, the probability that a ball is placed in them increases. So, a bin with a negative or zero $\hat{Gap}$ has a higher probability of a ball being allocated to it, whereby its estimated gap tends towards $0$ (in case of negative estimated gap-ed bins). On the other hand, bins with positive estimated gap receive a ball with low probability even when chosen as candidates, and their estimated gap decreases towards $0$. Hence, we observe that the estimated gap of any bin tends towards $0$. Since, estimated gap is the difference of the load and the estimated average of a bin and the gap tends to zero, the load of the bins becomes nearly equal to their estimated average w.h.p.
\end{proof}

\begin{theorem}
\label{th:constgap}
	The maximum load in any bin is $\lceil m/n \rceil + \Theta(1)$ w.h.p using the $IDEA$ allocation algorithm for the sequential, on-line and unweighted balls-into-bins problem.
\end{theorem}
\begin{proof}
	Using the above lemmas we observe that the estimated average of each bin finally becomes $\lceil m/n \rceil$ and the load in each bin is equal to its estimated 
	average w.h.p. Hence the maximum load in any bin is $\lceil m/n \rceil + \Theta(1)$ w.h.p.
\end{proof}

\begin{corollary}
\label{cor:load}
	The $IDEA$ algorithm provides a perfectly balanced allocation with constant gap.
\end{corollary}
\begin{proof}
	Since the maximum loaded bin has a load of $\lceil m/n \rceil + \Theta(1)$ w.h.p. (Theorem~\ref{th:constgap}), the \emph{Gap} is of $\Theta(1)$ providing a perfectly balanced allocation 
	for the balls-into-bins problem with constant gap.
\end{proof}

\section{Discussion}
\label{sec:disc}

We note that the \textit{Greedy[d]} algorithm can also retry $\gamma$ times to find a bin of even lower total number of balls that what it could do in a single round. Still, the distribution of the balls in bins will be different than the $IDEA$ algorithm because the $IDEA$ algorithm explicitly uses the \textit{expected gap} to make the decision of where the ball is placed. The key question is can the Greedy[d] algorithm give a constant gap and the answer is negative for a single retry because of the well known lower bound of $O(\ln\ln(n))$~\cite{azar}, while for multiple retries $\gamma$ has to be $\Theta(log(n))$~\cite{soda08} to achieve a constant gap. $IDEA$ however requires only constant ($< 2$) retries in the expectation (Lemma~\ref{lem:constgamma}), to achieve the constant gap. Further, it requires $\gamma = \frac{\phi}{d\epsilon}$ retries with high probability (Lemma~\ref{lem:negchoice}).

A bin, $B$ is chosen by $d$ balls among $n$ balls on expectation. However, the bin may be chosen $\alpha d$ times, $0 \leq \alpha \leq 1$ among the first $\rho$ balls that arrive. As such, the $Greedy[d]$ choice algorithm 
will place the balls in empty or lesser loaded bins if available. In the remaining balls, $B$ is chosen $(1-\alpha)d$ times. Now, for large values of $\alpha$, even if all these balls are placed in it, $B$ will have 
a load far less than the average of the system. So the $Gap$ increases. However, for $IDEA$ with large $\alpha$ values, the estimated average for $B$ will be large and hence its estimated gap will be significantly lower than 
the other bins. So, it has a higher probability of a ball being allocated to it. Thus, when the remaining balls arrive and a small fraction of them are placed in $B$, its load will still be closer to the actual average 
as compared to the d-choice algorithm. This sensitivity towards skewness in the random choices also enables $IDEA$ to arrive at a better allocation than the d-choice.

\section{Extended Framework}

\subsection{Weighted Case}
\label{subsec:ext_weight}

In this section we consider the weighted case of the balls-into-bins problem where the balls have weights drawn from a distribution $\chi$ with an expected weight $W^{*}$, such that the weight of any ball $W$ has a finite variance and can be bounded by $(W^*-k) \leq W \leq (W^*+k)$, where $k$ is a constant. We apply the $IDEA$ algorithm and show that the gap is also constant w.h.p. in such scenarios.

\begin{theorem}
\label{th:wconstgap}
	The maximum load in any bin is $W^*(\lceil m/n \rceil + \Theta(1))$ w.h.p using the $IDEA$ allocation algorithm for the sequential, on-line and weighted balls-into-bins problem.
\end{theorem}
\begin{proof}
	Reworking the lemmas stated in Section~\ref{sec:proof} we observe that the estimated average of each bin converges to $W^*\lceil m/n \rceil$ and that the load in each bin tends to its estimated 
	average w.h.p. Hence the maximum load in any bin is given by $W^*(\lceil m/n \rceil + \Theta(1))$ w.h.p. The complete proofs of the lemmas for the weighted case is provided in Appendix~\ref{sec:wproof}.
\end{proof}

\begin{corollary}
\label{cor:wload}
	The $IDEA$ algorithm provides a perfectly balanced weighted allocation with constant gap even for the general weighted case of the Balls-into-bins problem.
\end{corollary}
\begin{proof}
	From Theorem~\ref{th:wconstgap} we observe that as the maximum load is $W^*(\lceil m/n \rceil + \Theta(1))$. Hence $IDEA$ provides a perfectly balanced allocation for the 
	weighted case w.h.p. having a constant gap of $W^*\Theta(1)$.
\end{proof}

\subsection{Multi-Dimensional Case}
\label{subsec:ext_multi}

In this section, we consider the multidimensional (md), variant of the balls and bins problem. One multidimensional
variant, proposed by~\cite{multi} is as follows: Consider throwing $m$ balls into $n$ bins, where each ball is a 
uniform D-dimensional (0-1) vector of weight $f$. Here, each ball has exactly $f$ non-zero entries chosen uniformly 
among all $\binom{D}{f}$ possibilities. The average load in each dimension for each bin is given as $mf/nD$.

Let $l(a, b)$ be the load in the dimension $a$ for the $b^{th}$ bin. The gap in a dimension (across the bins) is given by
$gap(a) = max_b l(a, b) − avg(a)$, where $avg(a)$ is the average load in the dimension $a$. The maximum gap across all
the dimensions, $max_a gap(a)$, then determines the load balance across all the bins and the dimensions. Thus, for the
multidimensional balanced allocation problem, the objective is to minimize the maximum gap (across any dimension).
We refer to the multidimensional ball as md-ball and the multidimensional bin as md-bin.

In another variation of multidimensional balanced allocation the constraint of uniform distribution for populated
entries is removed. Here again, each ball is a D dimensional 0-1 vector and each ball has exactly $f$ populated dimensions,
but these populated dimensions can have an arbitrary distribution. In the third variation that is most general of
the three, the number of populated dimensions, $f$, may be different across the balls, where $f$ then is a random variable
with an appropriate distribution.

Each md-ball has $f$ populated dimensions, 
where $f$ could be constant across the balls or a random variable with a given distribution. Let, $s_i(t)$ denote the sum
of the loads (minus corresponding dimension averages) across all $D$ dimensions for the bin $i$ at time $t$, expressed as
$s_i(t) = \sum_{d=1}^D x^d_i$. This reduces the problem to that of the \emph{scalar weighted case}. The $IDEA$ algorithm works based 
on the sum of the dimensions for each bin. Also, for each choice of the bin, its estimated average is now incremented by $\frac{f}{d}$.

\begin{theorem}
\label{th:multi_constgap}
	For the multi-dimensional scenario, the $IDEA$ algorithm provides a constant gap for uniform distribution of the $f$ 
	populated dimensions for each ball with $m=n$.
\end{theorem}
\begin{proof}
Following the analysis in Section~\ref{subsec:ext_weight}, the $Gap$ in the system is bounded by $\Theta(1)$. Hence, the difference of the number of 
balls in the maximum bin and the actual average of the system is constant. For $m=n$, the average is $1$ and so the number of 
balls in the maximum bin is also a constant. Given a uniform distribution of the $f$ populated dimensions of each ball over $D$, 
the $Gap$ is bounded by $\Theta(1)$.
\end{proof}

\subsection{Parallel Case}
\label{subsec:parallel}

In this section we describe the algorithmic protocol to extend $IDEA$ for the parallel balls-into-bins scenario. In the parallel scenario multiple 
balls are allocated to bins simultaneously in a single round. The remain balls are considered for allocation in the next round. This process is 
repeated until all the balls are allocated. Later in this section we will show that the proposed protocol ensures that the algorithm completes 
in a finite number of rounds. We consider that in any round, $r$, a bin may accept only one ball.

Let $x$ balls be simultaneously allocated in round $r$. We observe that the outcome of round $r$ can be obtained by sequentially allocating $x$ 
balls by $IDEA$. Hence any round in the parallel case can be replaced by a series of sequential processes of $IDEA$. Hence the gap remains 
\emph{constant} even in the parallel case with $IDEA$. 

\begin{algorithm}[ht]
\begin{center}
	\caption{Communication Protocol}
	\label{algo:parallel}
	\begin{algorithmic}
		\REQUIRE Number of bins ($n$), Number of choices per ball ($d$)
		\ENSURE Parallel execution of $IDEA$

		\medskip
		
		\STATE Step 1. Each ball, $B_i$ chooses $d$ bins as candidates for allocation, and stores the choices as $M_i$.
		\STATE Step 2. Ball $B_i$ queries its chosen bins ($M_i$) for the \emph{estimated gap}.
		\STATE Step 3. The bins queries returns their estimated gap to the corresponding balls.
		\STATE Step 4. Ball $B_i$ selects the bin $b_i$ with the lowest estimated gap among its chosen bins and sends a confirmation message, $C1_i$.
		\STATE Step 5. A bin $b_j$ receiving a $C1_i$ message confirms allocation of ball $B_i$ and sends it a message $C2_{ij}$. If a bin receives multiple $C1_i$ messages, it arbitrarily selects one of them.
		\STATE Step 6. Ball $B_i$ after receiving $C2_{ij}$ sends message $INC$ to all its $d$ chosen bins ($M_i$) and commits to bin $b_j$.
		\STATE Step 7. All the bins in $M_i$ receiving $INC$ message increments their estimated average by $\frac{1}{d}$.
	\end{algorithmic}
\end{center}
\end{algorithm}

The communication protocol, as given in Algorithm~\ref{algo:parallel} ensures that there is no deadlock in the system and that each bin accepts 
at most one ball in each round. Since the allocation of a ball into a bin is done by \emph{two-way handshaking} between the 
ball and the bin, a bin may receive multiple confirmations from the balls but will accept only one of them, and since each ball makes a single 
choice of the bin where it prefers to be allocated, deadlock in the system is avoided. The update of the \emph{estimated average} of the bins receiving 
the $INC$ message is similar to that of the sequential $IDEA$ with the use of sampling.

We now prove that the algorithm terminates in finite number of rounds to guarantee a constant gap.

\begin{theorem}
\label{th:parallel}
	$IDEA$ in the parallel scenario using the communication protocol described in Algorithm~\ref{algo:parallel} provides a constant gap in 
	expected $O(\log \log n)$ rounds.
\end{theorem}
\begin{proof}
	Since each round of the parallel case of $IDEA$ can be simulated with multiple sequential processes of it, $IDEA$ along with the communication 
	protocol described above provides a constant gap.

	We observe that the execution of $IDEA$ is identical to that of the ordinary d-choice algorithm except for the parameter on which the allocations 
	of the balls are done. Hence Theorem 21 of~\cite{parallel} stating that the \emph{Threshold(1)} for parallel cases terminates after at most $\log \log n + O(1)$ 
	steps, holds in our case as well. However, each ball will select a bin zero or negative estimated gap in $\gamma$ retries. Hence the total number 
	of rounds taken by $IDEA$ in the parallel setting will be given by $\gamma \log \log n$. The expected value of $\gamma$ is a constant (Lemma~\ref{lem:constgamma}). 
	Hence the expected number of rounds for the algorithm to terminate is given by $O(\log \log n)$.
\end{proof}

It can easily been observed that this protocol still provides a constant gap even for the heavily loaded case when $m>>n$.

\section{Conclusions}
\label{sec:conc}

This paper proposes the \emph{Improved D-choice with Estimated Average}, $IDEA$ algorithm which w.h.p. provides a perfectly balanced allocation for the sequential, online 
and uniform weighted balls-into-bins problem. We propose a better metric for greedy placement of the balls using the estimated average of the system for each bin. 
We show that for a constant $d$ choice and expected constant number of rounds per ball, the maximum loaded 
bin in $IDEA$ is $\lceil m/n \rceil + \Theta(1)$ w.h.p. This result holds for $m=n$ case as well as the heavily loaded scenario where $m>>n$. We also extends the 
solution for the general weighted case (with $m>>n$) to show similar results for balls with weights taken from an arbitrary distribution with finite variance and 
for the multi-dimensional case with $m=n$ for uniform distribution of $f$ populated dimensions over the $D$ total dimensions. We also propose a communication 
protocol which in conjunction with $IDEA$ provides a constant gap with expected $O(\log \log n)$ rounds.

\pagebreak

{\small{
\bibliographystyle{abbrv}
\bibliography{stoc}
}}

\pagebreak

\appendix

\section{Sampling}
\label{sec:sampling}

Allocation of balls-into-bins for a single choice procedure has a \emph{Poisson} distribution approximately. We leverage this fact for the $d$ choice scenario to show that 
the sampling done by the $IDEA$ algorithm fairly accurately updates the estimated average of the bins w.h.p. 

Let $\lambda$ be the mean of the number of times a bin is chosen. Hence $\lambda = \frac{md}{n}$. Also assume the sample size to be $N$. Define $X$ 
to be the sum of the number of times the sampled bins to have been chosen. Since the number of times a bin is chosen is a random variable that follows Poisson's distribution (for a single choice process) 
and the choices of the bins are independent Poisson distributions each with mean $\lambda = dm/n$, the characteristics of the sample of size $N$, also follows a Poisson distribution with mean $N\lambda$. 
We would like $X$ to be bounded in the region $[\frac{N\lambda}{\beta}, \beta N \lambda]$ w.h.p., where $\beta$ is arbitrarily close to $1$. Applying Chernoff's bound we have,
\begin{align}
\label{eq:sampling}
	P(\frac{N\lambda}{\beta} \leq X \leq N\lambda \beta) = 1 - (P(X \leq \frac{N\lambda}{\beta}) + P(X\geq N\lambda \beta))
\end{align}
Given Poisson's tail bound,
\begin{align}
\label{eq:sampl_great}
	&P(X \geq N\lambda \beta) \leq \frac{e^{-N\lambda} \left(\lambda e\right)^{N\lambda \beta}}{\left(N\lambda \beta\right)^{N\lambda \beta}} 
	=\frac{e^{N\lambda(\beta - 1)}}{\beta ^ {N\lambda \beta}} = \left(\frac{e^{\beta-1}}{\beta^{\beta}}\right)^{N\lambda}
\end{align}
Substituting $\beta = 1 + \frac{1}{2^\omega}$ for some large $\omega > 1$, Eq.~\eqref{eq:sampl_great} becomes equal to $\left(\frac{e^{\frac{1}{2^\omega}}}{\left(1+\frac{1}{2^\omega}\right)^{1+\frac{1}{2^\omega}}}\right)^{N\lambda}$. 
Approximating $e^x$ to be less than $1 + x + x^2$ for small values of $x$, we observe that the above fraction is less than $1$. Replacing the fraction with $\frac{1}{\alpha}$, where $\alpha > 1$ and 
substituting it in Eq.~\eqref{eq:sampl_great} with the expected value of $\lambda$, we have,
\begin{align}
\label{eq:sampl_1/n}
	P(X \geq N\lambda \beta) \leq \left(\frac{1}{\alpha}\right)^{N. \frac{md}{n}} 
\end{align}
For $m>n\log n$, Eq.~\eqref{eq:sampl_1/n} becomes
\begin{align}
	P(X \geq N\lambda \beta) \leq \left(\frac{1}{\alpha^{\log n}}\right)^{Nd} = \frac{1}{\left(\alpha^{\frac{\log_{\alpha} n}{\log_{\alpha} e}}\right)^{Nd}} 
	\approx \frac{1}{n^{Nd+c}} \qquad \qquad \text{[where $c$ is a constant]} \nonumber
\end{align}
Hence, we observe that a \emph{constant} number of samples suffices to guarantee high probability for bounding $X$ within the factor of $\beta$ when $m>n \log n$. 
However when $m<n\log n$, we need $N=\log n$ samples for the same guarantee to hold. Similar results can thus be obtained for $P(X \leq \frac{N\lambda}{\beta})$. 
Hence Eq.~\eqref{eq:sampling} becomes,
\begin{align}
	P(\frac{N\lambda}{\beta} \leq X \leq N\lambda \beta) = 1 - (P(X \leq \frac{N\lambda}{\beta}) + P(X\geq N\lambda \beta)) \geq 1 - \frac{2}{n} \nonumber
\end{align}
Therefore, $IDEA$ needs to sample \emph{constant} or $\log n$ bins for the cases $m<n\log n$ or $m>n\log n$ respectively, for efficiently and accurately updating the 
estimated average of each bin to be close to that of the actual average of the system w.h.p.

We also calculate the total number of samplings (amount of communication) done by the $IDEA$ algorithm in the case $m < n\log n$. On arrival of $n$ balls, the expected number of times a 
bin is chosen is given by $d$. However, this is bounded by $d\log n$ w.h.p. A bin will sample $N$ other bins only when it is chosen more than $d$ times when $n$ balls have been thrown. Using the Poisson's tail bound, in the general case when $nk$ balls have been thrown ($k \in [1..\log n]$) the probability 
of a bin being chosen $\beta \lambda$ times ($\beta > 1$) is given by $\Pr_{(k)} = \frac{e^{-\lambda_k} \left(\lambda_k e\right)^{\beta \lambda_k}}{\left(\beta \lambda_k\right)^{\beta \lambda_k}}$, where $\lambda_{k}$ is the expected number of times a bin is chosen when $nk$ balls have been thrown.
Hence, the expected number of total samplings, $E[Samples]$ done when total $n \log n$ balls have been thrown is given by,
\begin{align}
	E[Samples] = \sum_{k=1}^{\log n} nd~Pr_{(k)} < nd \qquad \qquad \text{[By algebraic manipulations]} \nonumber
\end{align}
Since, $d$ is a constant, the expected number of samplings done by IDEA is $O(n)$ and the total communication done by $IDEA$ is less than that in the naive case when $d=\log n$. 

\section{Theoretical Framework for the Weighted Case}
\label{sec:wproof}

In this section, we provide a theoretical proof of the constant gap performance of the weighted version of the $IDEA$ algorithm. 
We follow the same proof sketch as in the case of ball with unit weight. Further, we too assume here $m$ balls and $n$ bins, $m \gg n$.

\begin{lemma}
\label{lem:wping}
	If each weighted ball chooses $d$ bins i.u.r. out of $n$ bins, each bin is chosen by $\frac{md}{n}$ balls on expectation, 
	and by at most $\frac{md}{n} \log n$ weighted balls with high probability.
\end{lemma}
\begin{proof}
	 Similar to Proof of Lemma \ref{lem:ping}.
\end{proof}

\begin{lemma}
\label{lem:wav}
	At any iteration, the estimated average of each bin is approximately equal to the current average w.h.p.
\end{lemma}
\begin{proof}
	We assume here that $Z$ balls have already arrived and have been placed among the $n$ bins. The number of balls that chose bin $B$ among its $d$ candidates is 
	$\frac{Zd}{n}$ on expectation, since each bin can be chosen by a ball with a probability of $\frac{d}{n}$. The number of such balls is also bounded 
	by $(1+\log n)\frac{Zd}{n}$ with high probability (by Lemma~\ref{lem:ping}). However, a bin does not increment its estimated average by more than $d$ 
	times when $n$ balls are thrown. For each selection of bin $B$, it increases its estimated average by $\frac{W}{d}$, which is bounded by $\frac{W^{*}-k}{d} \leq \frac{W}{d} \leq \frac{W^{*}+k}{d}$ . Hence the current value of $\hat{A_B}$ is given by,
	{\small{
	\begin{align}
		\hat{A_B} = \frac{Zd}{n} \cdot \frac{W^*\pm k}{d} = \frac{Z\left(W^*\pm k\right)}{n} \text{  , which is the current average.} \nonumber
	\end{align}
	}}
	Hence, the estimated average $\hat{A}$ of any bin is nearly equal to the actual average w.h.p.
\end{proof}

\begin{lemma}
\label{lem:wgap}
	The amortized sum of the estimated gap, $\hat{Gap}$ over all the bins is zero.
\end{lemma}
\begin{proof}
	Each ball chooses $d$ candidate bins uniformly and randomly and is finally allocated to the bin having the lowest estimated gap. Hence for all 
	the $d$ chosen bins, their estimated average increases by $W/d$. The load of $d-1$ bins which do not receive the ball remains same, 
	and thus their estimated gap decreases by the above factor. However, for the bin in which the ball is placed, its load increases by $1$ and its 
	estimated gap increases by $W\left(1-\frac{1}{d}\right)$. Applying the arguments presented in the proof of Lemma~\ref{lem:gap} and Cor.~\ref{cor:gap_zero}, 
	we observe that the sum of change of the estimated gap over the $d$ chosen bins in any iteration is $W\left(1-\frac{1}{d}\right) + (d-1)\frac{-W}{d} = 0$.
	Using similar analysis applied in the proof of Lemma \ref{lem:gap} it can be shown that the sum of the estimated gap is zero by amortized analysis.
\end{proof}

\begin{corollary}
\label{cor:wgap_zero}
The sum of the estimated gap over all bins is zero for arbitrary small number of balls allocated in the system.
\end{corollary}
\begin{proof}
Similar to Proof of Corollary~\ref{cor:gap_zero}.
\end{proof}

\begin{lemma}
\label{lem:wnegbins}
	The number of bins having a \emph{zero} or \emph{negative} estimated gap, $\hat{Gap}$ is $\Theta(n)$.
\end{lemma}
\begin{proof}
Using the arguments presented in the above lemmas, we provide a sketch of the proof below similar to that of Lemma~\ref{lem:negbins}.
Let there be $\alpha$ bins with positive $\hat{Gap}$, $\beta$ bins with negative estimated gap, and $\theta$ bins having $0$
estimated gap. Hence, $\alpha + \beta + \theta = n$. We would like to establish a lower bound on $\beta +\theta$.
In order to have minimum number of bins with negative or zero $\hat{Gap}$, the value of the gap should be minimum for bins with
a positive gap and maximum for bins with a negative gap. The minimum positive estimated gap for a bin is
$ZW_{min}-\frac{1}{d}\sum_{i=1}^{Z(d-1)}W_i \approx Z(W^*\pm k)\left(1 - \frac{d-1}{d}\right)$ when $Z(d-1)$ balls have arrived in the system, of which only $Z$ balls have been committed into the bin.
We have $W_{min} = min\{W_1, W_2, \ldots, W_{Z(d-1)}\}$.
The maximum negative estimated average that a bin may have in this case is $-\frac{\sum_{i=1}^{Z(d-1)}W_i}{d} \approx -\frac{Z(d-1)(W^* \pm k)}{d}$. Hence,
{\small{
\begin{align}
&\alpha.(Z(W^* \pm k)(1-\frac{d-1}{d})) + \beta.(-\frac{Z(W^* \pm k)(d-1)}{d}) + \theta .0 = 0 \qquad \qquad \text{[From Lemma~\ref{lem:gap}]} \nonumber \\
&\therefore \alpha = \beta(d-1) \nonumber
\end{align}
}}
Further, $\alpha + \beta + \theta = n$. Hence, d$\beta + \theta = n$. So, the number of bins with zero or negative $\hat{Gap}$ is $\Theta(n)$.

Arguing similarly in the lines of Corollary~\ref{cor:gap_zero}, we can claim that the gap is still $\Theta(n)$ even when each round has $f(n) = n^{\epsilon}$ balls, where $f(n)$ 
is not a constant.
\end{proof}

\begin{lemma}
\label{lem:wnegchoice}
	Each ball chooses at least one bin having negative estimated gap among its $d$ choices w.h.p. in $\gamma$ rounds.
\end{lemma}
\begin{proof}
	Similar to Proof of Lemma~\ref{lem:negchoice}.
\end{proof}

\begin{lemma}
\label{lem:wconstgamma}
	The expected number of rounds, $\gamma$ per ball to find a bin with zero or negative estimated gap is constant.
\end{lemma}
\begin{proof}
	Similar to Proof of Lemma~\ref{lem:constgamma}.
\rem{
	Let $p_i$ denote the probability that we find a zero or a negative bin at iteration $i$. Therefore, we have 
	{\small{
	\begin{align}
		p_i &= \left(\prod_{1}^{i-1} P_{pos}\right)\cdot P_{neg} = \prod_{1}^{i-1} \frac{1}{2^d} \cdot \left(1-\frac{1}{2^d}\right) = \frac{2^d - 1}{2^{id}} \nonumber
	\end{align}
	}}
	where $P_{pos}$ is the probability of selecting a bin with a positive gap and $P_{neg}$ is the probability of selecting a bin with a zero or negative gap.
	The expected number of rounds per ball, $\gamma$ to find a zero or a negative gap is given by,
	{\small{
	\begin{align}
	\label{eq:wexp}
		&E[\gamma] = \sum ip_{i} = \left(2^d-1\right)\sum_{i=1}^{\frac{\log{n}}{d}}\frac{i}{2^{id}}
	\end{align}
	Let,
	\begin{align}
	\label{eq:wsi}
		&S(i) = \sum_{i=1}^{\frac{\log{n}}{d}}\frac{i}{2^{id}} \\
	\label{eq:wsi+1}
		&\therefore \frac{S(i)}{2^d} = \sum_{i=1}^{\frac{\log{n}}{d}}\frac{i}{2^{(i+1)d}}
	\end{align}
	Subtracting Eq.~\eqref{eq:wsi+1} from Eq.~\eqref{eq:wsi}, we have
	\begin{align}
	\label{eq:wsum}
		&\left(1-\frac{1}{2^d}\right)S(i) = \frac{1}{2^d} - \frac{\log{n}}{d2^{1+\log{n}}} + \frac{1}{2^d\left(2^d-1\right)} \nonumber \\
		&\therefore S(i) \approx \frac{2^d}{\left(2^d-1\right)^2}
	\end{align}
	Substituting Eq.~\eqref{eq:wsum} in Eq.~\eqref{eq:wexp}, we have
	\begin{align}
	 	E[\gamma] &\approx 1 + \frac{1}{2^d-1}\\ \nonumber
	 	\Rightarrow E[\gamma] &< 2
	\end{align}
	}}
}
\end{proof}

\begin{lemma}
\label{lem:wload}
	The load of each bin tends to its estimated average.
\end{lemma}
\begin{proof}
	Similar to Proof of Lemma~\ref{lem:load}.
\rem{
	$IDEA$ places each ball into a bin with zero or negative $\hat{Gap}$, with high probability $1 - \frac{1}{n^\phi}$ (Lemma~\ref{lem:negchoice}) using $\gamma$ retries. When a ball is placed in a bin, its $\hat{Gap}$ increases. Thus, the probability that this bin will again get a ball lowers. On the other hand, the bins that had been chosen but the ball was not placed in them have a decrease in their estimated gap. Hence, the probability that a ball is placed in them increases. By this policy, the estimated gap of a bin tends to $0$. Since, estimated gap is the difference of the load and the estimated average of a bin and the gap tends to zero, the load of the bins tend to their estimated average w.h.p.
}
\end{proof}

\end{document}